\documentclass{article}

\usepackage{microtype}
\usepackage{graphicx}
\usepackage{subfigure}
\usepackage{booktabs} 

\usepackage[tbtags]{amsmath}
\usepackage{amssymb}
\usepackage{booktabs}
\usepackage[toc,page]{appendix}
\DeclareMathOperator{\Tr}{Tr}

\DeclareMathOperator{\sigmoid}{sigmoid}

\usepackage{hyperref}

\usepackage[accepted]{icml2020}

\icmltitlerunning{WaveNODE: A Continuous Normalizing Flow for Speech Synthesis}

\begin{document}

\twocolumn[
\icmltitle{WaveNODE: A Continuous Normalizing Flow for Speech Synthesis}

\icmlsetsymbol{equal}{*}

\begin{icmlauthorlist}
\icmlauthor{Hyeongju Kim}{to}
\icmlauthor{Hyeonseung Lee}{to}
\icmlauthor{Woo Hyun Kang}{to}
\icmlauthor{Sung Jun Cheon}{to}
\icmlauthor{Byoung Jin Choi}{to}
\icmlauthor{Nam Soo Kim}{to}
\end{icmlauthorlist}

\icmlaffiliation{to}{Department of Electrical and Computer Engineering and INMC, Seoul National University, South Korea}

\icmlcorrespondingauthor{Hyeongju Kim}{hjkim@hi.snu.ac.kr}

\icmlkeywords{Speech Synthesis, Generative Model, Flow, Neural Vocoder}

\vskip 0.3in
]

\printAffiliationsAndNotice{}

\begin{abstract}
In recent years, various flow-based generative models have been proposed to generate high-fidelity waveforms in real-time. However, these models require either a well-trained teacher network or a number of flow steps making them memory-inefficient. In this paper, we propose a novel generative model called WaveNODE which exploits a continuous normalizing flow for speech synthesis. Unlike the conventional models, WaveNODE places no constraint on the function used for flow operation, thus allowing the usage of more flexible and complex functions. Moreover, WaveNODE can be optimized to maximize the likelihood without requiring any teacher network or auxiliary loss terms. We experimentally show that WaveNODE achieves comparable performance with fewer parameters compared to the conventional flow-based vocoders.

\end{abstract}

\section{Introduction}
\label{intro}

\begin{figure*}[t]
	\centering
	\subfigure[]{
	\includegraphics[width=0.23\linewidth]{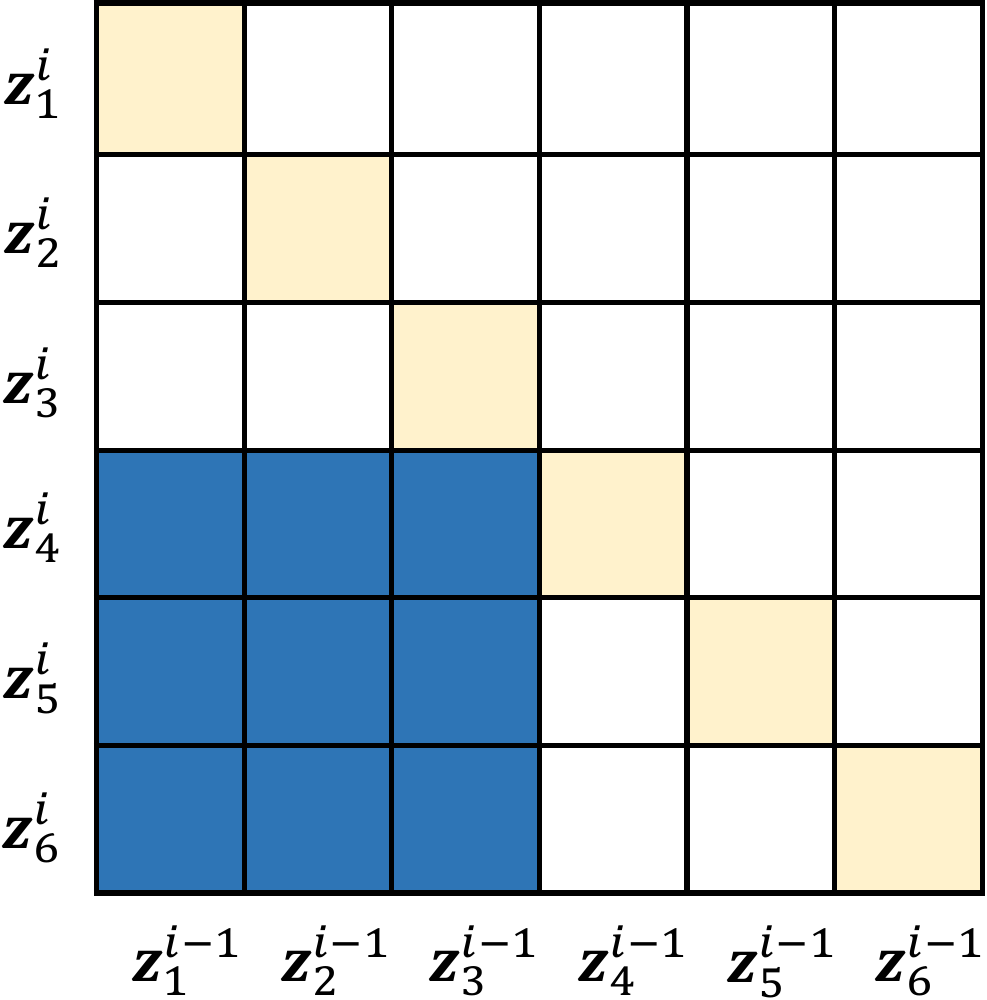}
	\label{fig:fig1a}
	}
	\hspace{0.05\textwidth}
	\centering
	\subfigure[]{
	\includegraphics[width=0.23\linewidth]{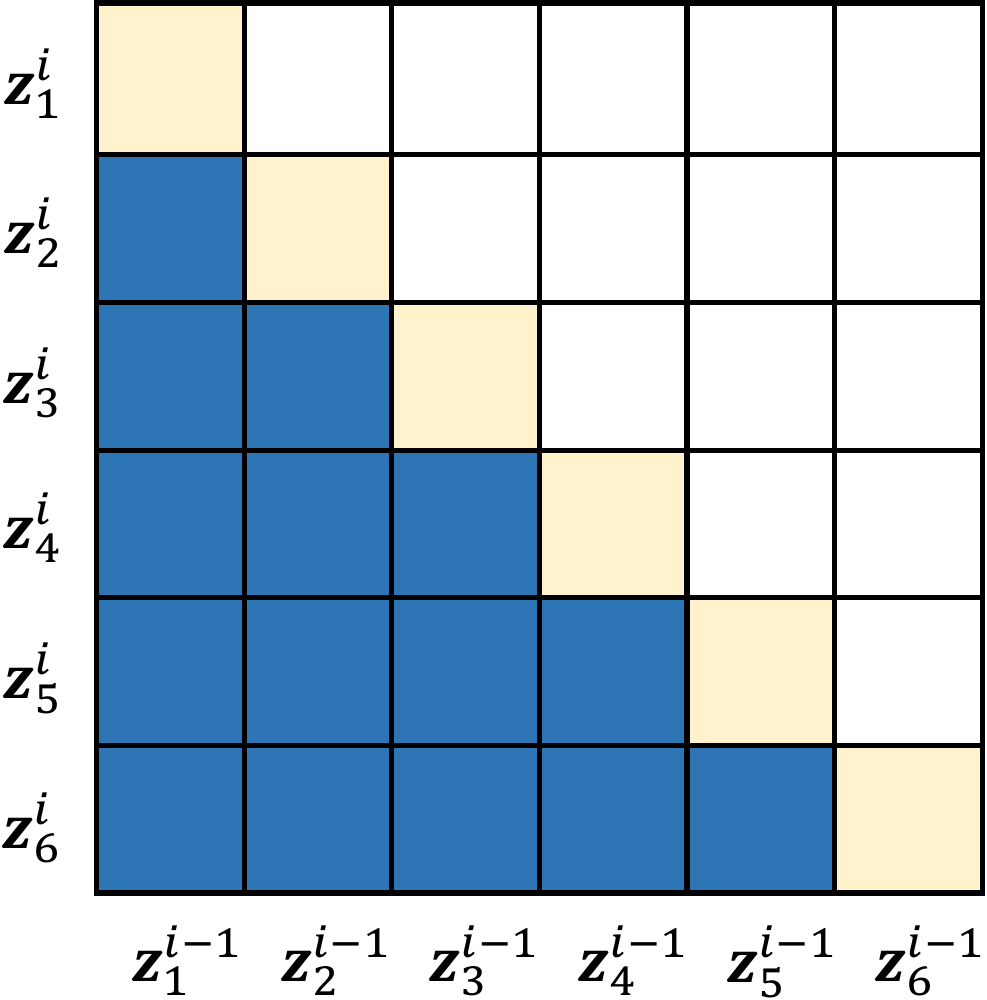}
	\label{fig:fig1b}
	}
	\hspace{0.05\textwidth}
	\centering
	\subfigure[]{
	\includegraphics[width=0.23\linewidth]{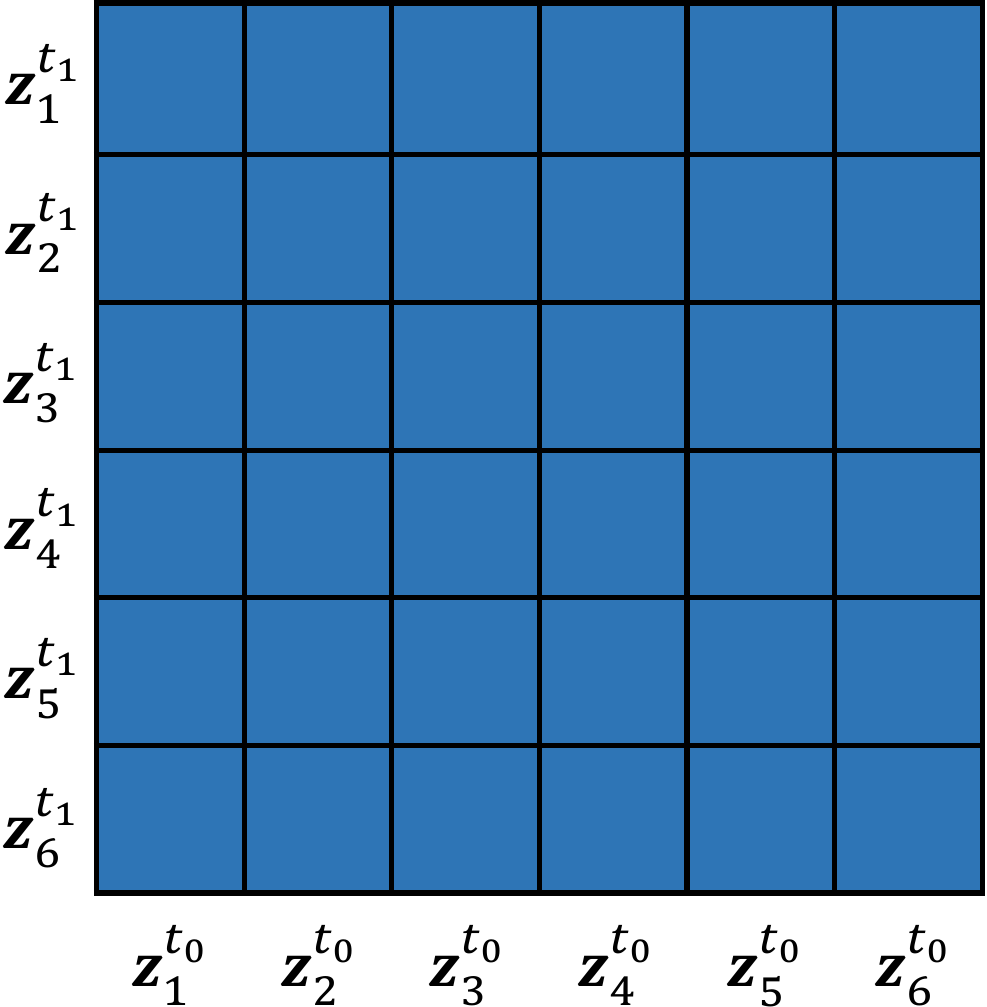}
	\label{fig:fig1c}
	}
    \caption{The Jacobian matrix of transformation in (a) Real NVP-based model, (b) IAF-based model, and (c) CNF-based model when the dimension of hidden state is 6. In (a) and (b), $\boldsymbol{z}^{i}_{d}$ stands for the $d$-th element of the $i$-th layer hidden state $\boldsymbol{z}^{i}$. In (c), $\boldsymbol{z}^{t}_{d}$ stands for the $d$-th element of time-varying variable $\boldsymbol{z}(t)$. The light-yellow cells represent the linear dependencies, the blue cells represent the non-linear dependencies and the white cells denote independent relations. The transformation in the CNF-based model is a lot more flexible than the others as the CNF does not impose any constraint on the type of functions used for flow transformation. 
    }
	\label{fig:jacobian}
\end{figure*}

Modern end-to-end speech synthesis models mostly consist of two stages: (1) transforming character embeddings to acoustic features such as mel-spectrograms~\cite{ping2017deep,shen2018natural,vasquez2019melnet,ren2019fastspeech}, and (2) synthesizing time-domain waveforms from the derived acoustic features~\cite{oord2016wavenet,oord2017parallel,ping2018clarinet,kim2018flowavenet,prenger2019waveglow}. In various end-to-end speech synthesis models, the WaveNet vocoder conditioned on mel-spectrograms is employed for the second stage to generate high-fidelity raw audio~\cite{ping2017deep,shen2018natural}. However, samples cannot be obtained in real-time with the WaveNet vocoder due to its autoregressive nature. To enable fast sampling, flow-based generative models have recently attracted attention in the field of speech synthesis~\cite{oord2017parallel,kim2018flowavenet,ping2018clarinet,prenger2019waveglow}.

In order to generate audio samples in real-time, parallel WaveNet~\cite{oord2017parallel} and ClariNet~\cite{ping2018clarinet} employ inverse autoregressive flow (IAF)~\cite{kingma2016improved} which takes advantage of the inverse transformation of an autoregressive function. Although IAF allows a parallel sampling procedure, it is not suitable to directly train the parallel WaveNet or ClariNet according to the maximum likelihood criterion. Instead, the parallel WaveNet and ClariNet are trained through probability density distillation which requires a well-trained teacher network. Additional hand-engineered objective functions are also needed to make the training procedure stable and to produce high quality audio. 

FloWaveNet~\cite{kim2018flowavenet} and WaveGlow~\cite{prenger2019waveglow} adopt an affine coupling layer which was originally proposed in Real NVP~\cite{dinh2016density}. The affine coupling layer provides a simple inverse transformation and tractable determinant of the Jacobian. Unlike the IAF-based flow models, both the inference and sampling processes are parallelizable so that these models can be trained according to the maximum likelihood criterion without any auxiliary loss terms. However, Real NVP-based models require a number of flow steps to perform density estimation accurately as the affine coupling layer is too inflexible and simple. In this respect, FloWaveNet and WaveGlow are considered inherently memory-inefficient models.

Recently,~\citet{chen2018neural} introduced a new technique in which the hidden units are assumed to be continuously time-varying. In this framework, the continuous-time dynamics of the hidden units and their probability densities are described by deep neural networks. A continuous normalizing flow (CNF) is specified by these two dynamics using the instantaneous change of variables formula. Contrary to the discrete normalizing flows, the CNF does not impose any restrictions on its architecture and allows to use a quite flexible function for flow transformation as shown in Fig.~\ref{fig:jacobian}. Here, in light of the advantages of the CNF, we propose a novel generative flow for speech synthesis called WaveNODE. WaveNODE generates high-fidelity waveforms from the corresponding mel-spectrograms with much fewer parameters compared to the conventional flow-based models by replacing the discrete normalizing flows with CNF.

\section{WaveNODE}
\label{section: sec3}
\begin{figure}[t]
	\centering
	\includegraphics[width=0.7\linewidth]{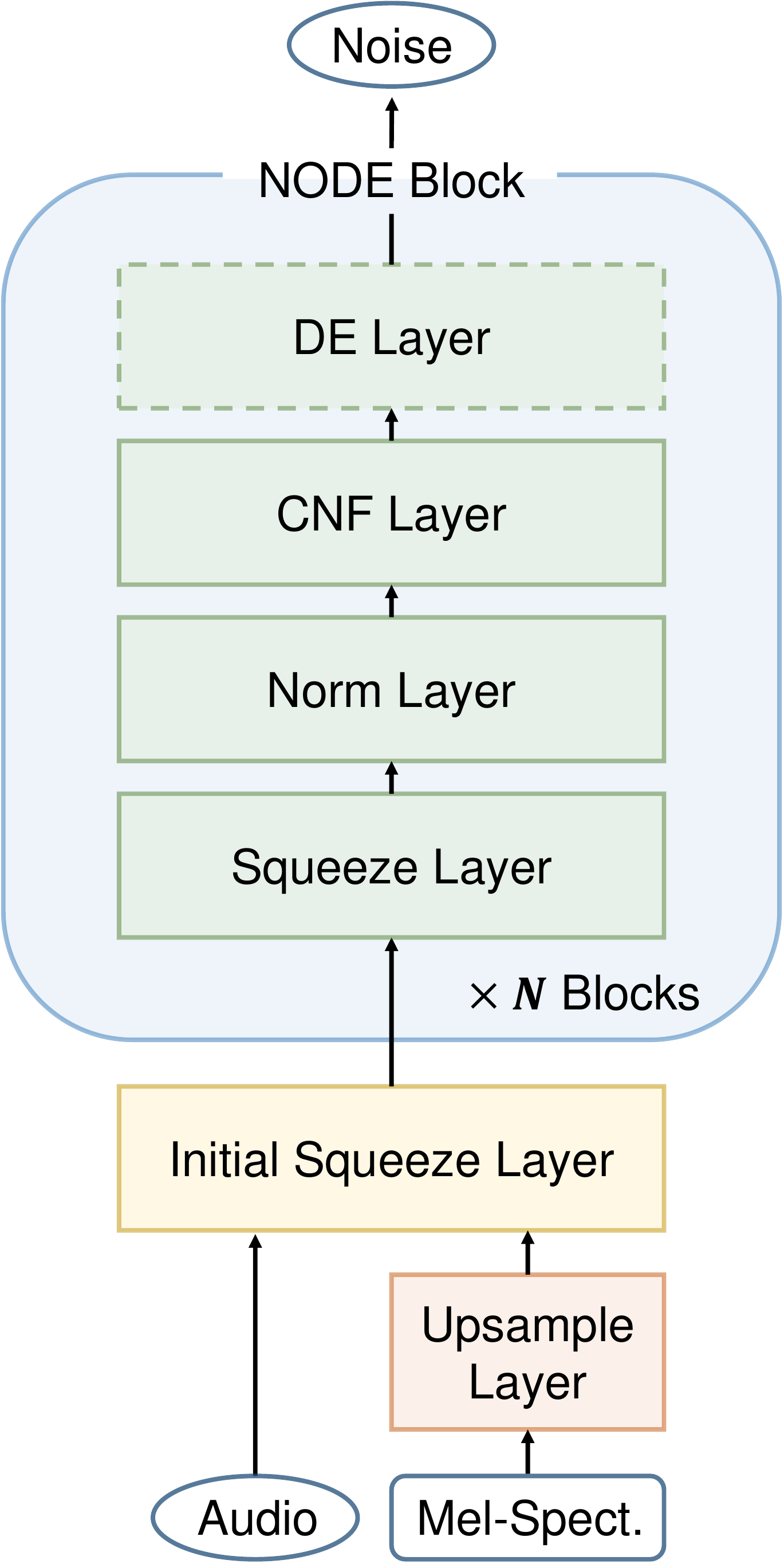}
    \caption{Overall structure of WaveNODE.}
	\label{fig:WaveNODE}
\end{figure}

We propose WaveNODE which takes advantage of the CNF for speech synthesis. WaveNODE is capable of generating high-fidelity waveforms from mel-spectrograms with a few flow steps. Moreover, WaveNODE does not require a teacher network or additional loss terms for training. The overall structure of the proposed WaveNODE is shown in Figure~\ref{fig:WaveNODE}. We describe the CNF and the hierarchical architecture of WaveNODE in the next subsections.

\subsection{Continuous Normalizing Flow}

In Neural ODEs~\cite{chen2018neural}, the continuous dynamics of time-varying hidden units $\boldsymbol{z}(t) \in \mathbb{R}^D$ are parameterized using an ODE
\begin{equation}
\label{eq:eq14}
\frac{d\boldsymbol{z}(t)}{dt} = \boldsymbol{f}(\boldsymbol{z}(t), t),
\end{equation}
where $\boldsymbol{f}(\boldsymbol{z}(t), t)$ is implemented by a neural network. Here, $t$ represents the time-step of solving the ODE, not the temporal axis of waveforms. Applying Eq.~\eqref{eq:eq14}, the change in log-density of $\boldsymbol{z}(t)$ follows a differential equation given by
\begin{equation}
\label{eq:eq15}
\frac{\partial \log p(\boldsymbol{z}(t))}{\partial t} = -\Tr\left(\frac{\partial \boldsymbol{f}(\boldsymbol{z}(t), t)}{\partial \boldsymbol{z}(t)}\right),
\end{equation}
which is called the instantaneous change of variables formula. To avoid the $\mathcal{O}(D^2)$ cost of computing the trace operation, an unbiased estimate of Eq.~\eqref{eq:eq15} can be derived using the Hutchinson's trace estimator~\cite{hutchinson1990stochastic} as follows:
\begin{equation}
\label{eq:eq16}
\begin{split}
\frac{\partial \log p(\boldsymbol{z}(t))}{\partial t} & = -\mathbb{E}_{p(\boldsymbol{\epsilon})} \left[\boldsymbol{\epsilon}^{\top} \frac{\partial \boldsymbol{f}(\boldsymbol{z}(t), t)}{\partial \boldsymbol{z}(t)} \boldsymbol{\epsilon}\right] \\
& \approx -\frac{1}{K} \sum_{k=1}^{K} \left[ \boldsymbol{\epsilon}_{k}^{\top} \frac{\partial \boldsymbol{f}(\boldsymbol{z}(t), t)}{\partial \boldsymbol{z}(t)} \boldsymbol{\epsilon}_{k} \right],
\end{split}
\end{equation}
where $\boldsymbol{\epsilon}_{k} \in \mathbb{R}^{D}$ is a noise vector drawn from $p(\boldsymbol{\epsilon})$ such that $\mathbb{E}[\boldsymbol{\epsilon}]=\boldsymbol{0}$ and $\mathrm{Cov}(\boldsymbol{\epsilon})=\boldsymbol{I}$~\cite{grathwohl2018ffjord}. Eq.~\eqref{eq:eq16} enables a fast computation of Eq.~\eqref{eq:eq15} since $\boldsymbol{\epsilon}_{k}^{\top} \frac{\partial \boldsymbol{f}(\boldsymbol{z}(t), t)}{\partial \boldsymbol{z}(t)}$ (vector-Jacobian product) can be efficiently calculated via deep learning libraries (e.g., PyTorch~\cite{paszke2019pytorch}) without explicitly writing out the Jacobian matrix. By setting ${K}=1$, we obtain an unbiased stochastic estimator of the dynamics of the log-likelihood with $\mathcal{O}(D)$ cost.

To train a generative model based on the CNF, we first assume that a latent variable $\boldsymbol{z}_{0} \in \mathbb{R}^{D}$ follows a simple distribution $p_{Z}(\boldsymbol{z}_{0})$. Next, let $\boldsymbol{f}(\boldsymbol{z}(t), t)$ be the dynamics of $\boldsymbol{z}(t)$ with the initial value $\boldsymbol{z}(t_{0})=\boldsymbol{z}_{0}$. Given a datapoint $\boldsymbol{x} \in \mathbb{R}^{D}$, the corresponding latent variable $\boldsymbol{z}_{0}$ is obtained by solving the following ODE:
\begin{equation}
\label{eq:eq17}
\boldsymbol{z}_{0} = \boldsymbol{z}(t_{0}) = \int_{t_{1}}^{t_{0}}\boldsymbol{f}(\boldsymbol{z}(t), t)dt + \boldsymbol{z}(t_{1}),
\end{equation}
where the final value $\boldsymbol{z}(t_{1}) = \boldsymbol{x}$. The log-likelihood of $\boldsymbol{x}$ can also be determined by solving another ODE
\begin{equation}
\label{eq:eq18}
\log p_{X}(\boldsymbol{x}) = \log p_{Z}(\boldsymbol{z}_{0}) - \int_{t_{0}}^{t_{1}} \boldsymbol{\epsilon}^{\top} \frac{\partial \boldsymbol{f}(\boldsymbol{z}(t), t)}{\partial \boldsymbol{z}(t)} \boldsymbol{\epsilon}dt,
\end{equation}
where $\boldsymbol{\epsilon}$ is a sampled noise vector from $p(\boldsymbol{\epsilon}$). Since $\boldsymbol{z}(t)$ varies along the vector field $\boldsymbol{f}(\boldsymbol{z}(t), t)$, the sampling process can be performed by simply reversing the time interval in Eq.~\eqref{eq:eq17} as follows:
\begin{equation}
\label{eq:eq19}
\boldsymbol{x} = \int_{t_{0}}^{t_{1}}\boldsymbol{f}(\boldsymbol{z}(t), t)dt + \boldsymbol{z}_{0}.
\end{equation}
Unlike the conventional normalizing flows, $\boldsymbol{f}(\boldsymbol{z}(t), t)$ is not required to be invertible or have a tractable Jacobian. Hence, any arbitrary functions can be employed for $\boldsymbol{f}(\boldsymbol{z}(t), t)$.

\subsection{CNF Layer}
Since we are interested in retrieving a time-domain signal from its mel-spectrogram, we apply a CNF framework to estimate the conditional distribution of audio samples. Given an upsampled mel-spectrogram $\boldsymbol{c}$ to full time resolution, Eq.~\eqref{eq:eq14} and Eq.~\eqref{eq:eq15} can be extended to a conditional formulation as follows:

\begin{equation}
\label{eq:eq20}
\frac{d\boldsymbol{z}(t)}{dt} = \boldsymbol{f}_{CNF}(\boldsymbol{z}(t), t, \boldsymbol{c}),
\end{equation}
\begin{equation}
\label{eq:eq21}
\frac{\partial \log p(\boldsymbol{z}(t)|\boldsymbol{c})}{\partial t} = -{\boldsymbol{\epsilon}}^{\top}\frac{\partial \boldsymbol{f}_{CNF}(\boldsymbol{z}(t), t, \boldsymbol{c})}{\partial \boldsymbol{z}(t)}{\boldsymbol{\epsilon}}.
\end{equation}
To capture the long-term dependency between audio samples, WaveNODE adopts a non-causal dilated convolutional network similar to the WaveGlow~\cite{prenger2019waveglow} architecture for $\boldsymbol{f}_{CNF}(\boldsymbol{z}(t), t, \boldsymbol{c})$. Let $V$ and $W$ be a convolutional layer, and $U$ be a linear projection. The activation function of the layers used for $\boldsymbol{f}_{CNF}(\boldsymbol{z}(t), t, \boldsymbol{c})$ is defined as
\begin{equation}
\label{eq:eq22}
\boldsymbol{a}^{f}_{in} = W_{f} * \boldsymbol{z}(t) + V_{f} * \boldsymbol{c} + U_{f} t,
\end{equation}
\begin{equation}
\label{eq:eq23}
\boldsymbol{a}^{g}_{in} = W_{g} * \boldsymbol{z}(t) + V_{g} * \boldsymbol{c} + U_{g} t,
\end{equation}
\begin{equation}
\label{eq:eq24}
\boldsymbol{a}_{out} = \tanh (\boldsymbol{a}^{f}_{in}) \odot \; \sigmoid (\boldsymbol{a}^{g}_{in}),
\end{equation}
where * denotes a convolution operator, super/subscripts $f$ and $g$ denote filter and gate, respectively. Note that WaveNODE uses $t$ as a global condition via broadcasting and $\boldsymbol{c}$ as a local condition. The CNF layer receives the initial value $\boldsymbol{z}(t_{0})$, time interval $[t_{0}, t_{1}]$ and the condition $\boldsymbol{c}$ as inputs. Using a black-box ODE solver, the CNF layer outputs the final value $\boldsymbol{z}(t_{1})$ and the change in log-likelihood $\Delta \log p(\boldsymbol{z}(t))$.

\subsection{Squeeze Layer}
\label{squeeze}

A squeeze layer rearranges an input tensor of shape ($C \times L$) to form an output tensor of shape ($qC \times \frac{L}{q}$) where $q$ is a scale factor. Increasing the number of channels by $q$ times, the squeeze layer enlarges the receptive field of the convolutional networks linearly. This also helps each NODE block to focus on different temporal dependencies.

Since speech signals have a very high temporal resolution, it may not be desirable to directly use a mono audio signal $\boldsymbol{x}$ with the shape ($1 \times L$) as input to a convolutional network. To deal with this, WaveNODE employs the initial squeeze layer to transform an input tensor of shape ($1 \times L$) into an output tensor of shape ($q_{init} \times \frac{L}{q_{init}}$) given an initial scale factor $q_{init}$. This is the same as using Squeeze Layer with a scale factor $q \cdot q_{init}$ in the first NODE Block.

\subsection{Norm Layer}

WaveNODE incorporates a norm layer to alleviate the difficulties that arise when training deep neural networks. In this work, we consider two variants of batch normalization~\cite{ioffe2015batch}.

\subsubsection{Actnorm}

With trainable parameters $\boldsymbol{s}$ and $\boldsymbol{b}$ ($\boldsymbol{s}, \boldsymbol{b} \in \mathbb{R}^{C}$), actnorm~\cite{kingma2018glow} performs per-channel affine transformation on $\boldsymbol{z}(t) \in \mathbb{R}^{C \times L}$
\begin{equation}
\label{eq:eq25}
\boldsymbol{f}_{AN}(\boldsymbol{z}(t)_{c}) = \boldsymbol{s}_{c} \cdot \boldsymbol{z}(t)_{c} + \boldsymbol{b}_{c},
\end{equation}
where $\boldsymbol{s}_{c}$ and $\boldsymbol{b}_{c}$ are the $c$-th elements of $\boldsymbol{s}$ and $\boldsymbol{b}$, respectively, and $\boldsymbol{z}(t)_{c} \in \mathbb{R}^{L}$ is the $c$-th channel vector of $\boldsymbol{z}(t)$. The parameters $\boldsymbol{s}$ and $\boldsymbol{b}$ are initialized to normalize the pre-actnorm activations given an arbitrary initial batch (i.e., data dependent initialization). The change in log-likelihood obtained by passing $\boldsymbol{z}(t)$ through the actnorm layer can be computed as
\begin{equation}
\label{eq:eq26}
\Delta \log p(\boldsymbol{z}(t)) = -L \sum_{c=1}^{C} \log |\boldsymbol{s}_{c}|.
\end{equation}

\subsubsection{Moving Batch Normalization}

Moving batch normalization (MBN) exploits running averages instead of the current batch statistics as given by
\begin{equation}
\label{eq:eq27}
\boldsymbol{f}_{MBN}(\boldsymbol{z}(t)_{c}) = \boldsymbol{s}_{c} \cdot \left(\frac{\boldsymbol{z}(t)_{c}-\boldsymbol{\mu}_{c}}{\boldsymbol{\sigma}_{c}}\right) + \boldsymbol{b}_{c},
\end{equation}
where $\boldsymbol{\mu}$ and $\boldsymbol{\sigma}$ are the running averages of mean and standard deviation for each channel ($\boldsymbol{\mu}, \boldsymbol{\sigma} \in \mathbb{R}^{C}$), and subscript $c$ denote the $c$-th channel component of variables. Similar to Eq.~\eqref{eq:eq26}, the change in log-likelihood can be computed as
\begin{equation}
\label{eq:eq28}
\Delta \log p(\boldsymbol{z}(t)) = -L \sum_{c=1}^{C} \left(\log |\boldsymbol{s}_{c}| -\log |\boldsymbol{\sigma}_{c}|\right).
\end{equation}

\subsection{NODE Block}

The NODE block is the primary component of WaveNODE, which basically consists of a squeeze layer, a norm layer, and a CNF layer as shown in Figure~\ref{fig:WaveNODE}. Similar to FloWaveNet~\cite{kim2018flowavenet}, WaveNODE stacks several NODE blocks and factors out half of the feature channels at a selected few NODE blocks. Factored-out channels are assumed to be Gaussian whose mean and variance are computed via a density estimation layer (DE layer) using the remaining channels as inputs. These statistics are used to evaluate the likelihood of the factored-out channels. The remaining channels are further passed through deeper NODE blocks and transformed into standard Gaussian noise in the end. 

\section{Experiments}
\label{section: sec4}

In order to evaluate the performance of WaveNODE, we conducted a set of experiments using the LJ speech dataset~\cite{ljspeech17} with the Griffin-Lim algorithm~\cite{griffin1984signal} and various neural vocoders. All the neural vocoder models were trained for 7 days on a single NVIDIA RTX 2080Ti GPU. For the subjective evaluation of audio fidelity, we performed a 5-scale mean opinion score (MOS) test with 33 audio examples per model and 27 participants. The audio examples were randomly selected from the test set. Each participant listened to the audio examples played in random order and evaluated the audio quality. Confidence intervals of MOS were calculated using the method proposed in~\citet{ribeiro2011crowdmos}. To encourage reproducibility, we attach the code for WaveNODE and the audio samples used in the experiments
\footnote{\url{https://github.com/ANLGBOY/WaveNODE/}.}
\footnote{\url{https://wavenode-example.github.io/}.}. Also, we describe the configuration of other vocoders in Appendix~\ref{appendixA}.

\subsection{WaveNODE}

WaveNODE has 4 NODE blocks, each of which basically consists of a CNF layer, an actnorm layer, and a squeeze layer with scale factor $q$ = 2. For the CNF layer, WaveNODE employs a 4-layer non-causal WaveNet with kernel size 3 where the channels of residual and skip connections are set to 128. Since WaveNODE stacks only a few NODE blocks, we set the base of dilation to 3 to increase the receptive field. At the end of the second NODE block, half of feature channels are factored out and their likelihood is estimated via a DE layer with a 2-layer network. The initial scale factor $q_{init}$ is set to 4. For upsampling mel-spectrograms, a single layer of transposed 1D convolution is incorporated.


\section{Results}
\label{section: sec5}

\subsection{Audio Fidelity and  Conditional Log-Likelihood}

\begin{table}[t]
\caption{Mean opinion scores (MOS) with 95\% confidence intervals and conditional log-likelihoods (CLL) on the test set. The CLL obtained by solving neural ODEs is labeled with $\dagger$.}
\label{table:model-comparison}
\begin{center}
\begin{small}
\begin{sc}
\begin{tabular}{cccc}
\toprule
Model        & \begin{tabular}[c]{c}Number of\\ parameters\end{tabular} & CLL & MOS        \\ \midrule
Ground Truth & -                                                              & -        & 4.84$\pm$0.06\\
Griffin-Lim  & -                                                              & -        & 2.82$\pm$0.26 \\
WaveNet      & 4.8M                                                           & 4.616    & 4.48$\pm$0.16 \\
WaveGlow     & 87.9M                                                          & 4.501    & 4.17$\pm$0.15 \\
WaveGlow     & 17.1M                                                          & 4.366    & 1.75$\pm$0.20 \\
FloWaveNet   & 182.6M                                                         & 4.449    & 2.99$\pm$0.19 \\
FloWaveNet   & 18.6M                                                          & 4.249    & 1.22$\pm$0.18 \\
WaveNODE     & 16.2M                                                          & 4.497$^\dagger$    & 3.53$\pm$0.18 \\ \bottomrule
\end{tabular}
\end{sc}
\end{small}
\end{center}
\vskip -0.1in
\end{table}

\begin{figure*}[ht]
\vskip 0.15in
\begin{center}
\centerline{\includegraphics[width=2\columnwidth]{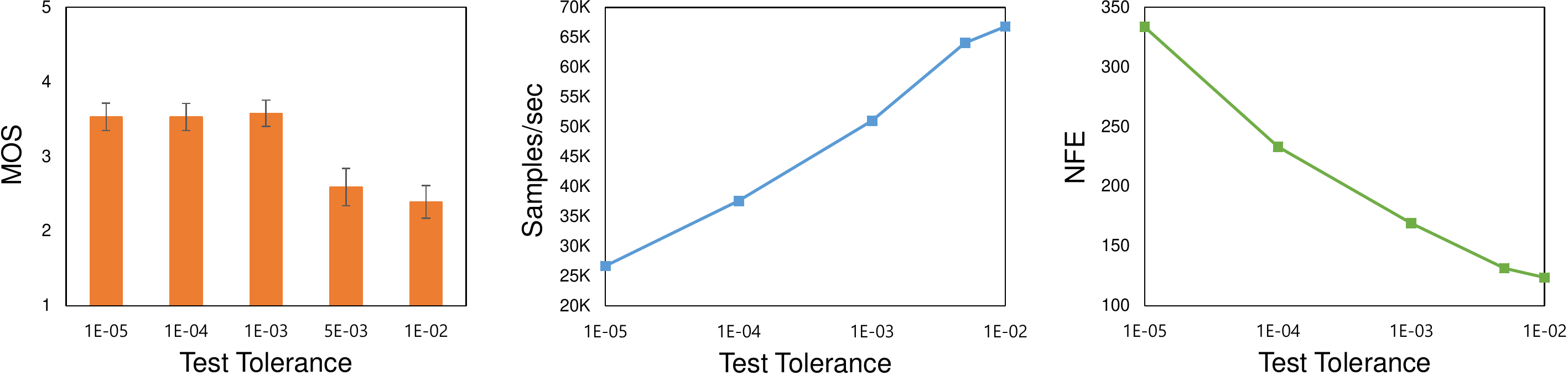}}
\caption{Mean opinion scores (MOS) with 95\% confidence intervals (left), the number of samples generated per second (middle) and the number of function evaluations (NFE) (right) versus test tolerance. The results demonstrate that we can boost the sampling speed of WaveNODE by lowering the tolerance to some extent without significantly degrading audio fidelity.}
\label{fig:test-tolerance}
\end{center}
\vskip -0.2in
\end{figure*}

We report the results of model comparison on a 5-scale MOS and conditional log-likelihoods (CLL) in Table~\ref{table:model-comparison}. In both the MOS and the CLL tests, the performance of the WaveNet vocoder was the best among all vocoders used in the experiments. Among the flow-based vocoder models, WaveGlow gained the highest scores for both MOS and CLL of 4.17 and 4.501, respectively. The MOS score of WaveNODE was between the scores of WaveGlow and FloWaveNet. Note that WaveGlow and FloWaveNet have a much larger number of parameters than WaveNODE. To verify that only WaveNODE is capable of generating high-fidelity audio with a few flow steps, we also evaluated the performance of the compressed models of WaveGlow and FloWaveNet. As shown in Table~\ref{table:model-comparison}, the compact models of WaveGlow and FloWaveNet received relatively poor MOS scores of 1.75 and 1.22, respectively. The results demonstrate that the ability of WaveNODE to use constraint-free functions for flow transformation allows the implementation of a memory-efficient vocoder that is capable of generating high-fidelity waveforms.

\subsection{Synthesis Speed}

\begin{table}[t]
\caption{Comparison of synthesis speed. All models are benchmarked using a single RTX 2080Ti GPU.}
\label{table:sample-speed}
\vskip 0.15in
\begin{center}
\begin{small}
\begin{sc}
\begin{tabular}{cc}
\toprule
Model      & Samples/sec \\ \midrule
WaveNet    & 56          \\
WaveGlow   & 328,690     \\
FloWaveNet & 320,062     \\
WaveNODE   & 51,045      \\ \bottomrule
\end{tabular}
\end{sc}
\end{small}
\end{center}
\vskip -0.1in
\end{table}

WaveNODE is able to control the synthesis speed by tuning the accuracy of the black-box ODE solver. When solving an ODE, the ODE solver predicts the errors and adjusts its step-size to reduce the errors below a user set tolerance. To study the effect of changing the accuracy of the ODE solver, we tested the synthesis speed on a single RTX 2080Ti GPU. More specifically, we divided the total number of generated sample points by the total time, measured the number of function evaluations (NFE), and evaluated the audio quality by modifying the tolerance which had been fixed to $10^{-5}$ during training. The middle and right graphs in Figure~\ref{fig:test-tolerance} represent that the synthesis speed increased steadily by allowing higher tolerances at logarithmic scale. On the other hand, the audio quality was little affected until the tolerance was set to $10^{-3}$. However, the audio quality dropped sharply after setting the tolerance to higher than $5\times10^{-3}$. The results suggest that WaveNODE can increase the sampling speed to some extent without seriously degrading the audio fidelity.

To compare the synthesis speed of the various neural vocoders, we counted the number of samples generated per second and report the results in Table~\ref{table:sample-speed}. We set the test tolerance of WaveNODE to $10^{-3}$. While WaveNet achieved the best result in the MOS test, it showed the worst performance on synthesis speed as it generates one sample point at a time (i.e., ancestral sampling). On the other hand, WaveNODE generated 51K samples/sec, which was a lot faster than WaveNet due to the parallel sampling process of flow operation. This suggests that WaveNODE is capable of generating audio samples in real-time even though WaveNODE has to solve complex ODEs in every flow operation. The sampling speeds of FloWaveNet and WaveGlow were the fastest since these models are not required to solve ODEs.

\subsection{Type of Norm Layer}

\begin{table}[t]
\caption{Evaluations on mean opinion score (MOS) and conditional log-likelihood (CLL) for WaveNODE models with different types of norm layer.}
\label{table:norm-layer}
\vskip 0.15in
\begin{center}
\begin{small}
\begin{sc}
\begin{tabular}{cccc}
\toprule
Model    & Norm layer & CLL      & MOS        \\ \midrule
WaveNODE & Actnorm    & 4.497    & 3.53$\pm$0.18 \\
WaveNODE & MBN        & 4.460    & 3.36$\pm$0.17 \\
WaveNODE & None       & 4.457    & 3.22$\pm$0.17 \\ \bottomrule
\end{tabular}
\end{sc}
\end{small}
\end{center}
\vskip -0.1in
\end{table}

\begin{table*}[t]
\caption{Training configurations of flow-based vocoder models.}
\label{table:limitation}
\vskip 0.15in
\begin{center}
\begin{small}
\begin{sc}
\begin{tabular}{ccccccc}
\toprule
Model      & Batch size & Epoch & Iteration & CLL & MOS        & Training time \\ \midrule
WaveGlow   & 8          & 240   & 354K      & 4.501    & 4.17$\pm$0.15 & 7 days        \\
FloWaveNet & 2          & 138   & 814K      & 4.449    & 2.99$\pm$0.19 & 7 days        \\
WaveNODE   & 20         & 46    & 27K       & 4.497    & 3.53$\pm$0.18 & 7 days        \\
WaveNODE   & 4          & 26    & 77K       & 4.452    & 2.91$\pm$0.17 & 7 days        \\ \bottomrule
\end{tabular}
\end{sc}
\end{small}
\end{center}
\vskip -0.1in
\end{table*}

In Table~\ref{table:norm-layer}, we report the performance of WaveNODE models with different norm layers. The results show that the norm layer is advantageous for WaveNODE to achieve good performance when adopting a multi-scale architecture unlike WaveGlow which can be fully trained without batch normalization. Popular CNF-based models often employ an MBN layer for stable training while constructing a deep architecture~\cite{grathwohl2018ffjord,yang2019pointflow}. In our experiments, WaveNODE showed the finest results in terms of both MOS and CLL when employing an actnorm layer rather than the MBN layer. This implies that CNF-based models for speech data can be trained more efficiently by exploiting the actnorm layer.

\subsection{Analysis of Training Progress}

\begin{figure}[t]
\vskip 0.1in
	\centering
	\includegraphics[width=0.84\linewidth]{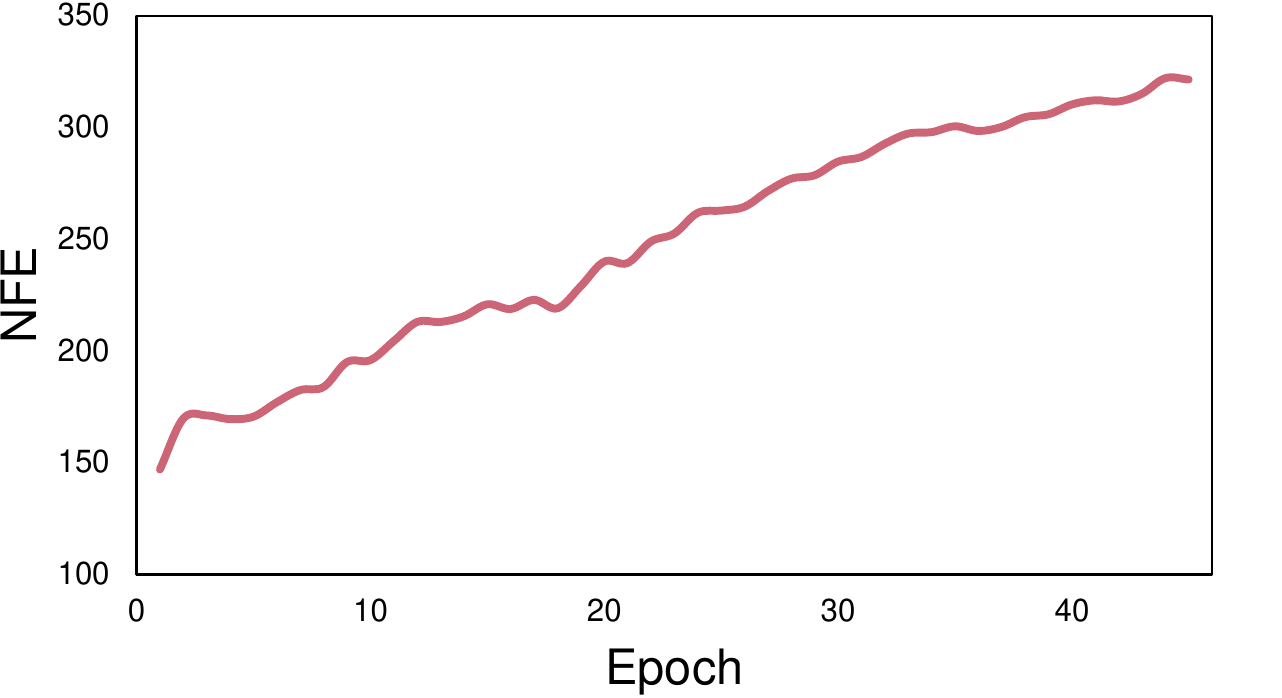}
    \caption{Evolution of the number of function evaluations (NFE) during training.}
	\label{fig:Limitation}
\vskip -0.1in
\end{figure}

One of the major drawbacks shared by CNF-based models is the computational cost of the black-box ODE solver. The ODE solver creates a deep computational graph since it finds the solutions to complex ODEs specified by neural networks in an iterative way. Due to this large amount of computation, the mini-batches in the CNF-based models are processed for a long time. Table~\ref{table:limitation} shows that processing time per mini-batch in WaveNODE was significantly longer than the conventional flow-based models. While WaveGlow went through 240 epochs in 7 days, WaveNODE performed only 46 epochs during the same period. Interestingly, on the other hand, WaveNODE showed decent performance with less training steps. This implies that WaveNODE is trained more efficiently per mini-batch due to the flexible functions used for flow transformation. We also tested whether reducing the batch size to increase the number of iterations can improve the performance of WaveNODE, but the resultant performance was degraded as shown in Table~\ref{table:limitation}.

It has been reported that the NFE in CNF-based models increases as training progresses~\cite{chen2018neural,grathwohl2018ffjord}. We also observed a similar phenomenon when training WaveNODE and report the overall trend of the NFE consumed for inference in Figure~\ref{fig:Limitation}. The main reason for this phenomenon is that ODEs become more complicated to accurately estimate the conditional distribution of waveforms. Since NFE directly affects the time taken to process a mini-batch, we plan to research how to prevent an increase of NFE in future work.

\subsection{Analysis of Dilation}
\begin{table}[t]
\caption{Comparison of WaveNODE models with different dilations.}
\label{table:dilation}
\vskip 0.1in
\begin{center}
\begin{small}
\begin{sc}
\begin{tabular}{cccc}
\toprule
Model    & \begin{tabular}[c]{c}Dilation\\ (at $i$-th layer)\end{tabular} & CLL   & MOS  \\ \midrule
WaveNODE & $3^{i}$                                                               & 4.497 & 3.53$\pm$0.18 \\
WaveNODE & $2^{i}$                                                               & 4.408 & 3.17$\pm$0.17 \\ \bottomrule
\end{tabular}
\end{sc}
\end{small}
\end{center}
\vskip -0.1in
\end{table}

WaveNODE is composed of only a few flow steps unlike the conventional flow-based models, which might result in the small receptive field. In order to capture the long-range temporal dependencies in audio signals, we basically set the base of dilation to 3 in WaveNODE for the previous experiments. To verify the effect of dilation, we trained the WaveNODE model with dilation of $2^{i}$ at the $i$-th layer and evaluated performance in terms of MOS and CLL. Indeed, the dilation was critical to the quality of audio samples generated by WaveNODE as shown in Table~\ref{table:dilation}. We found that WaveNODE produces a trembling sound when the dilation is a multiple of 2 due to the narrow receptive field.

\section{Conclusion}
\label{section: sec6}
In this work, we presented the novel generative model, namely WaveNODE, which leverages a CNF for speech synthesis.
We successfully applied the CNF framework to a large-dimensional data (e.g., audio) without any additional loss term. In the experiments, we demonstrated that WaveNODE shows comparable performance with fewer flow steps compared to the conventional flow-based models. Also, we verified that WaveNODE is able to synthesize audio samples in real-time due to the parallel sampling process. We believe that applying CNF to speech synthesis can be further developed and refined to produce more realistic waveforms.

\section*{Acknowledgments}

This work was supported by Samsung Research Funding Center of Samsung Electronics under Project Number SRFC-IT1701-04.

\nocite{langley00}

\bibliography{WaveNODE}
\bibliographystyle{icml2020}

\clearpage

\appendix
\section{Model Configuration}
\label{appendixA}
\subsection{Griffin-Lim}

The Griffin-Lim algorithm~\cite{griffin1984signal} estimates the signal from its modified short-time Fourier transform (STFT) magnitude in an iterative way. For the experiments, we first approximated the STFT magnitude from the mel-spectrogram and applied the Griffin-Lim algorithm with 32 iterations for time-domain conversion.

\subsection{WaveNet}
\label{WaveNet}

We trained an autoregressive WaveNet whose output is a single Gaussian distribution. It has been shown that a single Gaussian WaveNet is capable of modeling raw waveforms without degradation compared to WaveNet models with mixture distribution~\cite{ping2018clarinet}. We stacked 2 dilated residual blocks of 10 layers with kernel size 2 and set the number of hidden units in both residual and skip connections to 128. To upsample the mel-spectrograms from frame-level to sample-level resolution, we employed two layers of transposed 2D convolution with one leaky ReLU activation.

\subsection{FloWaveNet}

FloWaveNet~\citep{kim2018flowavenet} consists of 8 context blocks, each of which contains 6 flow operations. For the affine coupling layers, FloWaveNet employs a 2-layer non-causal WaveNet with kernel size 3. FloWaveNet uses 256 channels for residual and skip connections. Also, FloWaveNet factors out half of the feature channels after 4 context blocks and uses another 2-layer network to estimate the distribution of factored-out channels. FloWaveNet incorporates the same upsampling module described in Section~\ref{WaveNet}. 

For the experiments, we trained the compact version of FloWaveNet as well as the original model. We used 4 context blocks with 3 flow operations and factored out half of the feature channels after 2 context blocks for the compact model.

\subsection{WaveGlow}

WaveGlow~\cite{prenger2019waveglow} is composed of 12 blocks, each of which contains an affine coupling layer and an invertible 1x1 convolution. WaveGlow employs 8-layer non-causal WaveNet networks for the affine coupling layers and one layer of transposed 1D convolution for upsampling mel-spectrograms. WaveGlow factors out 2 of feature channels after every 4 blocks.

In addition to the original WaveGlow network, we also trained the compressed model for the experiments. We stacked 9 blocks, used  4-layer networks, and factored out 2 of feature channels after every 3 blocks for the compressed model.


\end{document}